# Synchronization in Quantum Key Distribution Systems

Anton Pljonkin [1,*], Konstantin Rumyantsev [1] and Pradeep Kumar Singh [2]

[1] Institute of computer technology and information security, Southern federal university, 347900, Taganrog, Russia
[2] Department of CSE, Jaypee University of IT, 173234, Waknaghat, Solan, HP, India
* Correspondence: pljonkin@mail.ru; Tel.: +7-905-459-2158

**Abstract:** In the description of quantum key distribution systems, much attention is paid to the operation of quantum cryptography protocols. The main problem is the insufficient study of the synchronization process of quantum key distribution systems. This paper contains a general description of quantum cryptography principles. A two-line fiber-optic quantum key distribution system with phase coding of photon states in transceiver and coding station synchronization mode was examined. A quantum key distribution system was built on the basis of the scheme with automatic compensation of polarization mode distortions. Single-photon avalanche diodes were used as optical radiation detecting devices. It was estimated how the parameters used in quantum key distribution systems of optical detectors affect the detection of the time frame with attenuated optical pulse in synchronization mode with respect to its probabilistic and time-domain characteristics. A design method was given for the process that detects the time frame that includes an optical pulse during synchronization. This paper describes the main quantum communication channel attack methods by removing a portion of optical emission. This paper describes the developed synchronization algorithm that takes into account the time required to restore the photodetector's operation state after the photon has been registered during synchronization. The computer simulation results of the developed synchronization algorithm were analyzed. The efficiency of the developed algorithm with respect to synchronization process protection from unauthorized gathering of optical emission is demonstrated herein.

## 1. Introduction

The main problem associated with confidential information transmission is distribution of the secret key between correspondents. Achieving the ultimate confidentiality during transmission of messages is only possible if the problem of key distribution is solved. Security of existing telecommunication networks is limited by computing means of the intruder. A physical solution of the key distribution problem is known as quantum cryptography, and it is based on the coding of a single particle quantum state. The main point of quantum cryptography is reliable distribution of a single key between legitimate users. The idea of quantum cryptography is attractive, as it allows for the creation of an ultimately random secret key. Its confidentiality and inability of monitoring by any unauthorized person is based on the laws of quantum physics [1,2–7]. Traditional cryptography methods are based on mathematical patterns, so there is a potential for their decoding. To ensure the ultimate security of a cryptographic scheme, certain conditions are to be fulfilled: the key must be absolutely random, its length must be greater or equal to the length of the encoded message, and the key is to be a one-off type.

Quantum cryptography is based on the following statements: it is impossible to clone an unknown quantum state, and it is impossible to extract the information on non-orthogonal quantum states without a disturbance. Hence, any measurement performed by an intruder will cause a change in the information carrier quantum state.

## 2. Quantum Key Distribution

Implementation of quantum cryptography is based on quantum key distribution systems (QKDSs). Fiber-optic systems using the phase coding of photon states, which are built according to a scheme with automatic compensation of polarization distortions, have an obvious advantage among the successfully implemented commercial examples of quantum key distribution (QKD) systems [8]. Such QKDSs have proven to be reliable systems when operating in key data generation modes. A QKDS with automatic compensation includes two stations interconnected by means of a fiber-optic line (quantum channel). A block diagram of the optical part of the QKDS is shown in Figure 1.

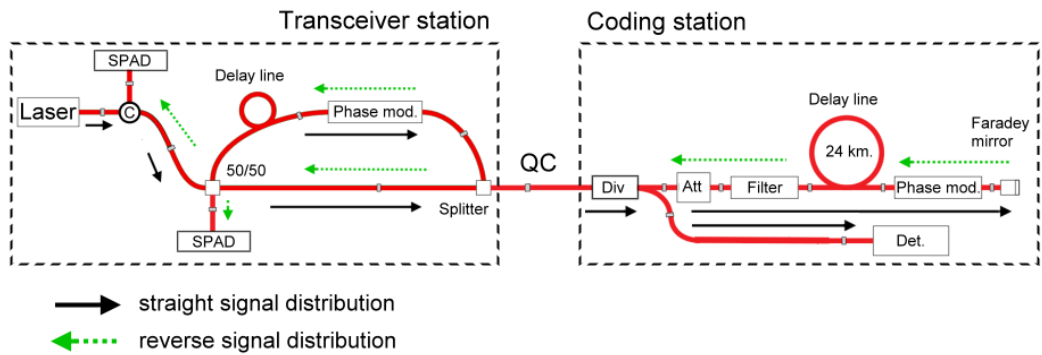

**Figure 1.** Optical scheme of the QKD system. SPADs—single photon avalanche photodiodes; 50/50, Div—optical dividers; Splitter—Beam splitter; QC—quantum channel (fiber-optic line between stations); Att—optical attenuator; Det—classical detectors.

Some papers [8–11] provide a detailed description of fiber-optic self-compensation quantum key distribution systems with phase coding of photon states. Note that QKD security is based not on mathematical methods and algorithms, but on the laws of quantum physics. In this case, protocols of quantum cryptography operate with optical signals attenuated to the photon level. Therefore, the average number of photons in the optical pulse is up to 0.1. The latter is to be understood as the availability of the signal only in each 10th generated pulse, but not as a photon quality factor.

## 3. Synchronization in a QKDS

To ensure functioning of the QKDS system, some preparatory procedures are required. In this manner, generation and distribution of quantum keys in such systems are always preceded by configuring and synchronization of spatially distributed stations. A QKDS does not function without initial synchronization, a process during which the key parameters of detection equipment are determined. Synchronization in self-compensating QKDSs consists in the high-accuracy measurement of the path length and propagation path of the optical emission from transceiver station to encoding station and vice versa. In this case, optical signals propagate in both directions in the same fiber-optic line [12–17]. It should be noted that, when measuring the total propagation length, one must consider not only the length of quantum channel between two stations but the length of each fiber-optic component inside the stations as well (Figure 1). For registering the attenuated optic emission, single-photon avalanche diodes (SPADs) were used in QKDS. When the cryptographic protocol is used, SPADs operate in single photon counting mode (Geiger mode). In QKDSs, using the scheme as shown above, the synchronization is based on the registration of receiving the optical pulse by photodetectors. During synchronization, the laser diode of the transceiver station generates a periodic optical pulse sequence. In this case, the clock pulses themselves act as time markers.

Measuring the propagation length consists in breaking the time interval equal to the pulse-repetition period into subintervals. When describing the synchronization, we use the notion of a time frame, which corresponds to the optical pulse repetition period. During synchronization, the time frame is divided into time segments, i.e. time windows. In fact, optical pulse detection comes, as a result, to detection of the time window that contains the optical pulse. During synchronization, each time window is analyzed in sequence. During the analysis, conversion of a photon into a primary electron (photoelectron, PE) or the receiving of dark current pulses (DCPs) is registered. When the entire time frame is analyzed, the time window that has the largest number of triggering (for both PE and DCP) is considered as a signal window, and the other time windows are considered as noise windows (those that do not contain synchronization pulses). A detailed description of the synchronization process is described in [15,17].

Results of actual tests of the self-compensating quantum key distribution system with phase coding of photon states are given in [11,18]. During experimental tests, it has been established that the synchronization in a QKDS is realized in a linear mode. Linear (multiphoton) mode is characterized by a large average number of photons in the pulse, which amounts to hundreds or thousands. It is known that a powerful optical signal, during transmission, is potentially vulnerable to unauthorized data gathering by diverting a portion of the optical emission from the communication channel. Due to the implementation of the multiphoton mode during synchronization, it is potentially easier for an attacker to get unauthorized access to the parameters of a quantum communication channel [12,13]. Algorithms monitoring the transmitted signals do not function at the stage of QKDS synchronization. Hence, a minor power loss of optical signal will not disrupt system operation and will not allow the intruder to have access into the quantum communication channel. Physically, a portion of optical power can be withdrawn without much effort with the use of fiber optic directional couplers or special "clips" [18,19]. The latter determines whether it is relevant to develop the synchronization algorithms that provide increased QKDS security.

In [17,19], an algorithm detecting the optical pulse for QKDS was developed. A distinct feature of the algorithm is that it uses attenuated optical pulses as synchronization signal. The average number of photons in the clock pulse is not more than 0.5. Photon registration is ensured by means of SPADs.

*Specifications of SPAD*

It is known that SPAD characteristics differ from the characteristics of an ideal single-photon photodetector. First, during the time window analysis, SPAD registers only the first received photon. It must be noted that analysis time during the synchronization process refers to the time when SPAD operates in photon counting mode, which is equal to the time window length. Second, when a photon or DCP is registered, a certain amount of time is required for SPAD to restore the operating condition. The latter parameter is commonly considered as the SPAD "dead time."

Consider the algorithm that detects an optical pulse during synchronizing QKDS. A laser diode generates optical pulses that have a wavelength of 1550 nm and a duration about 1 ns. It is assumed that the repetition period $\Delta T_s$ and a duration of the photon pulse $\Delta \tau_s$ are absolutely stable. It is necessary to note that the repetition period $\Delta T_s$ is determined by the length of quantum communication channel between QKDS stations. The time frame, which is equal to the optical pulse repetition period $\Delta T_s$ is divided into time windows $N_w$, whose length is $\tau_w$ such that $T_s = N_w \tau_w$. Photodetectors are switched into operation mode, and a sequential examination of all time windows begins. Each window is analyzed N times, where N is the sampling scope. When analyzing each time window, the number of registered PEs and/or DCPs is recorded. Thus, when all the time windows $N_w$ have been examined, an array of registered PE and/or DCP values is being formed. When a photon pulse is registered entirely in a single time window, correct detection is possible if two conditions are met. First, at least one PE or DCP must be registered in the signal time window during the analysis. Second, a number of pulses registered in the signal window must strictly exceed the number of pulses registered in all other noise windows. When a photon pulse is distributed between two adjacent windows, correct detection is possible if the other conditions are met:

(1) At least one PE or DCP must be registered in one of the two windows containing a photon pulse.

(2) The number of registered pulses in the first signal time window must strictly exceed the number of registered pulses in the second signal window and the number of generated DCPs in all remaining noise windows.

(3) The number of registered pulses in the second signal window that contains a portion of the photon pulse shall strictly exceed the number of registered pulses in the first signal window and the number of generated DCPs in all remaining noise windows.

(4) If the number of pulses accumulated in two adjacent time windows is equal, there is a decision that a photon pulse is received by any of these windows, if the number of accumulated pulses in the window exceeds the number of pulses registered in the remaining windows.

## 4. Development of the Synchronization Process

The index of optical emission refraction in an optical fiber core is known ($n_{fiber}$ = 1.49). The optical signals propagation speed for a fiber-optic line is calculated as

$$v_{fiber} = c_{opt}/n_{fiber} = 300000/1.49 = 201000 \text{ km/s}$$

where $c_{opt}$ is the speed of optical waves in vacuum.

The length of the fiber-optic line between two QKDS stations, $L_{FOL}$, can reach 100 km. Taking into consideration the back propagation in a two-way self-compensating fiber-optic system with phase coding of photon states, the length of optical pulse repetition period shall exceed $T_{s.min} = 2 \times L_{FOL}/v_{fiber} \approx 1$ ms in order to exclude counter pulses overlapping at $L_{FOL}$ = 100 km. Thus, the maximum optical pulse repetition frequency in a fiber-optic line with the length of 100 km shall not exceed $f_{s.max} = 1/T_{s.min} \approx 1$ kHz. The length of an optical pulse is considered to be $\tau_s = 1$ ns, and it is selected based on the parameters applicable to QKDS with laser emission sources. In [15], it was established that the length of the time window applicable in the detection process is to be chosen on the basis of the following condition:

$$\tau_w = (2 \ldots 4) \times \tau_s.$$

It has been determined that a single-photon photodetector registers all incoming photons, instantly recovering its operational state. Due to the latter, a consistent interrogation of time windows within a time frame is possible. In addition, another condition shall be noted, which greatly simplifies the implementation of digital equipment. Events that shall be counted during synchronization must be a multiple of 2.

Let us assume that the length of the fiber-optic line between two QKDS stations is $L_{FOL}$ = 100 km. For single-mode optical fiber Corning®SMF-28e+ at an operating wavelength of 1550 nm. The refraction index of optical emission in the core is $n_{fiber}$ = 1.4670. Thus, emission propagation speed in the core of a single-mode optical fiber is $v_{fiber} = c_{opt}/n_{fiber} \approx 205000$ km/s. Taking into consideration the backward propagation, an optical pulse repetition period shall exceed $T_{s.min} = 2L_{FOL}/v_{fiber} \approx 978$ mks, and maximum optical pulse repetition frequency shall not exceed 1 kHz to exclude counter pulses overlapping. The transmitting optical module generates optical clock pulses whose length is $\tau_s = 1\ ns$ and whose repetition period is $T_s = 1\ ms$ ($f_s = 1/T_s \approx 1$ kHz). Thus, the duration of the time window is $\tau_w = 2\tau_s$ = 2 ns. The number of time windows required for examination is $N_w = T_s/\tau_w = 5 \times 10^5 = 500000$. The nearest number that satisfies the multiple of 2 condition at $T_s \geq 1$ ms is 524288 = 2^19. This allows for more specific requirements of the repetition period of the generated optical clock pulses $T_s = N_w \times \tau_w = 524288 \times 2 = 1048576$ ns ≈ 1.05 ms ($f_s = 1/T_s \approx 954$ Hz). The increase of the time frame does not exceed 5%. If the taken sample scope is equal to N = 256 = 2^8, then the total time required for analysis of 256 time frames with the use of ideal single-photon photodetector is 256 × 1.05 = 268.8 ms.

In the developed algorithm, the average number of photons included in the pulse does not exceed 0.5. In a fiber-optic line whose length is 100 km, losses in the optical fiber make up 20 dB, i.e., the signal will be attenuated 100 times. Thus, the average number of PE will be only 0.005. The latter must be taken into consideration during synchronization subsystem development. Additionally, when performing the calculations, it is necessary to consider the SPAD "dead time" parameter, $\tau_{dead}$, whose typical value is 50 ns (average). This indicates that a certain time $\tau_w - \tau_{dead}$ is required to restore the SPAD operating conditions after the time window has been analyzed. The latter requires the notion of module $\tau_m$, whose duration cannot be less than the dead time.

Thus, it is reasonable to consider an error that increases module duration in relation to the duration of the time window. In this manner, for example, $d_{time} = 100$ μs ensures maximum photodiode quantum efficiency and minimum DCP frequency. On this basis, the system can analyze several temporary windows within a single time frame. Analysis time is not affected by the availability of the registered signal in the time window. Detection time delay (time required for SPAD to switch to Geiger mode) at each optical pulse sending is calculated by the following formula:

$$Z_t = \frac{T_s}{4} \times (A_n - 1) + \tau_w \times (B_n - 1)$$

where $A_n$ is the sequence number of SPAD activation in the frame; $B_n$ is the pulse (frame) sequence number.

It must be mentioned that SPAD time parameters were examined to estimate their influence on the synchronization efficiency only in QKDSs with phase coding of photon states.

## 5. Modeling

It is established that the position of the time window containing the signal pulse does not affect the probabilistic characteristics of the synchronization. The average number of signal PE in the pulse, DCP frequency, and sample scope in each time window have a significant effect on the probability of signal time window detection. Initial simulation data correspond to actual QKDS parameters. Constant parameters are an optical pulse length of 1 ns, a time window length of 2 ns, a pulse repetition period of 1048576 ns, and a number of analyzed time windows of 524288. The variables are the DCP frequency, average number of PEs in an attenuated optical pulse, and the sampling scope in each time window.

Figures 2–5 show the relation between the probability of correct detection of the time window containing a photon pulse and the DCP frequency.

Dependency diagrams shown in Figures 2 and 3 were built in each time window for a given sample scope, which corresponds to the values 32, 64, 128, 256, 512, 1024, and 2048. In Figure 2, we can see that the maximum probability of correct detection reaches 99.68% at average PE number of 0.01. Here, the highest probability is ensured at a DCP frequency of 25 Hz and a sample scope of 1024. At the same time, a change in DCP frequency to 400 Hz helps to reduce the detection probability to 95.65%, with a similar sample scope, which is more than 4%. We shall note the dependence type for the sample scope of 2048, in this case the difference in limiting probability values is more than 15%, and the decrease is proportional to the increase in DCP frequency. The latter is caused by the increase of average DCP number in the time window when the number of "views" of the time window with fixed PE value is increased. The dependencies presented in Figure 3 are of the greatest interest. With an average PE number of 0.5, one can allocate a sample range where the optimal probability of correct detection is ensured. It is evident that the best results are shown by the dependencies with the sample scope of up to 256. For the latter, the difference in detection probability limits does not exceed 1% (99.99% and 99.72%).

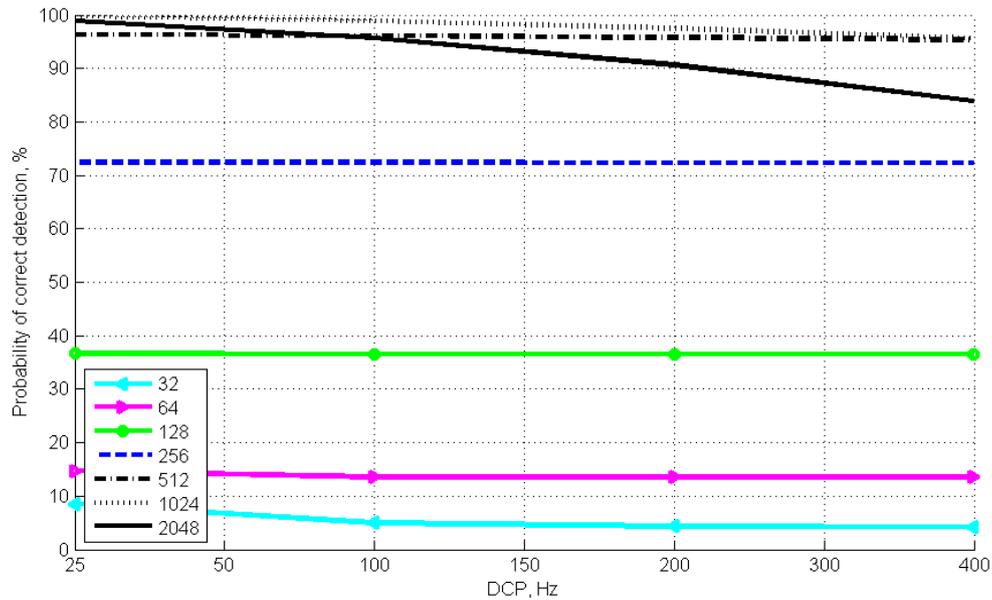

**Figure 2.** Probability of correct detection of a signal time window when the average number of photoelectrons (PEs) in a pulse is 0.01.

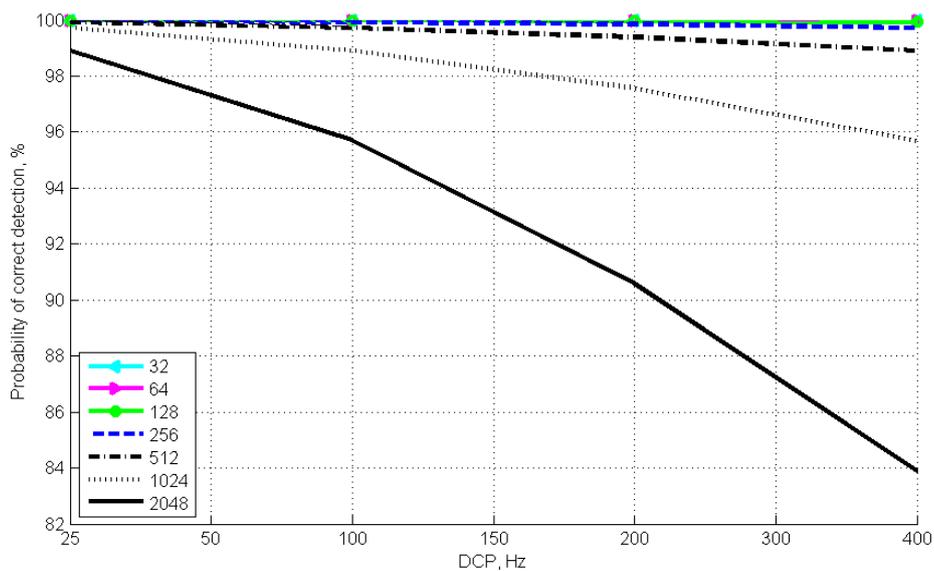

**Figure 3.** Probability of correct detection of a signal time window when the average number of PEs in a pulse is 0.5.

It shall be noted that the average PE number and the DCP frequency are not specified by the QKDS operator and therefore are not monitored. In contrast to the latter, the sample scope can be configured according to the requirements of the synchronization algorithm.

Figures 4 and 5 show the relation between the probabilities of time window detection and the sample scope average PE number and the DCP frequency are varying.

Figure 4 shows that the change in DCP frequency still does not have a considerable effect on the probability calculation, if the sample size is less than 1024. However, the growth dynamics can be traced right from the first values of the sampling scope. If the sampling scope is 128, the probability of detection is 36%, and if its value is 512, detection probability reaches 95%. We can come to conclusion that for the given relation of detection equipment parameters (Figure 4) an optimal sampling scope is a value that shall be within the range from 256 to 512. This allows for a detection probability from 95.6% to 99.68%.

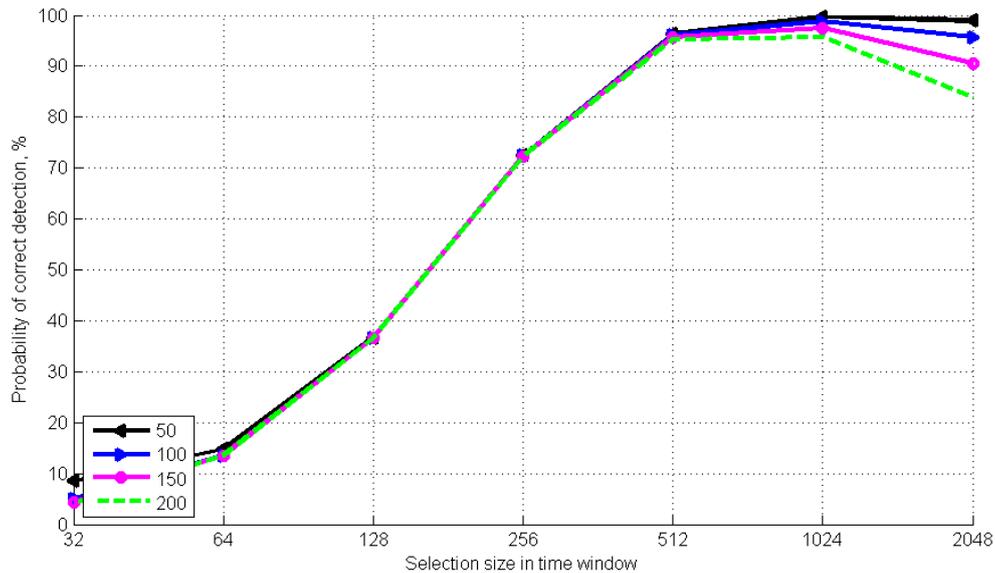

**Figure 4.** Probability of correct detection of a signal time window when average number of PEs in a pulse is 0.01.

Figure 5 shows the relation between the probability of time window correct detection and the sampling scope in the time window when the average number of PEs in the attenuated optical pulse is 0.5. We shall note that this value is the upper limit of the photon pulse parameter in the developed synchronization algorithm. Figure 5 shows that, if the sampling scope is less than 512, and the DCP frequency is of 50 to 200 Hz, the probability of correct detection does not fall below 98.9%. The simulation results determine the optimal sampling scope for the calculations. In such a manner, if the DCP frequency is less than 200 Hz, the best probability (99.92%) is ensured at a sampling scope less than 128. On the other hand, for the sampling scope of 256, the probability is 99.72%. Using a SPAD with a lower DCP frequency (<200 Hz) will ensure a detection probability above 99.97%, while the probability of erroneous detection is less than 0.03%.

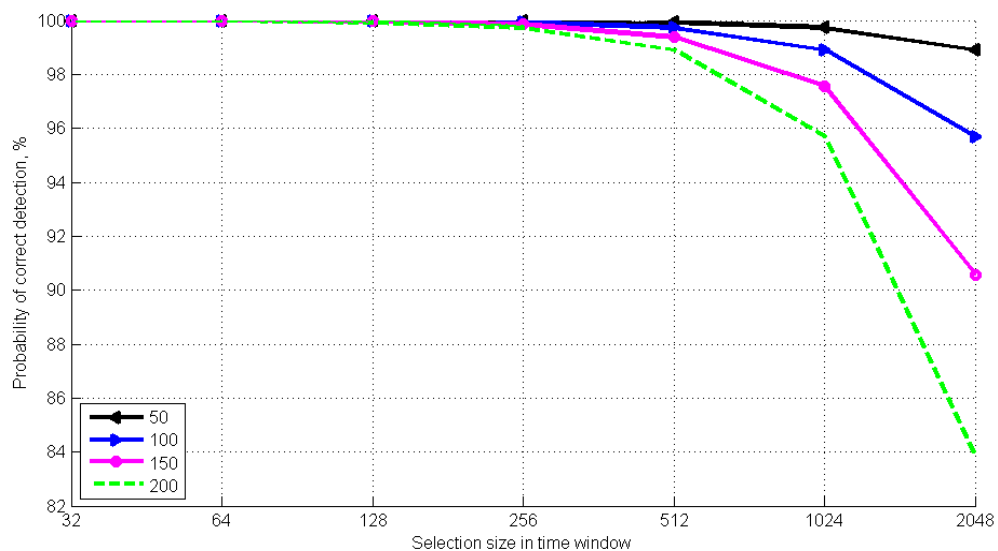

**Figure 5.** Probability of correct detection of a signal time window when average number of PEs in a pulse is 0.5.

Thus, simulation of the developed algorithm proves the efficiency of using attenuated optical pulses as synchronization signals. If an average photon number in an optical clock pulse is less than 0.5, increased security while establishing communication with a QKDS is ensured.

## 6. Conclusions

An algorithm that detects the time window with a clock pulse is herein described. A specific feature of the examined algorithm is that it is realized in a single-photon regime, where the average number of photons in an attenuated optical pulse does not exceed 0.5. The latter provides increased security of the QKDS synchronization mode. The process design method that detects the period that includes an optical pulse during synchronization and that considers the specifications of the SPAD applied in the QKDS is described. The computer simulation results of the time interval detection process are analyzed. Optimal criteria for limiting the sampling scope in each time window to ensure a probability of correct detection higher than 99.9% were established.

**Acknowledgments:** The study was supported by International Pljonkins corporation and Institute of computer technology and information security.

**Author Contributions:** Pljonkin A. developed an algorithm, analyzed data, and performed the experiments; Rumyantsev K. developed the idea and analyzed the experimental data; Pradeep K. checked the data and wrote the paper.

**Conflicts of Interest:** The authors declare no conflict of interest.